\documentclass[prl,aps,twocolumn,nofootinbib,floatfix]{revtex4}
\usepackage{graphicx}
\usepackage{bm}
\usepackage{amssymb}

\def\be{\begin{equation}}
\def\ee{\end{equation}}
\def\ba{\begin{eqnarray}}
\def\ea{\end{eqnarray}}
\def\half{\frac{1}{2}}
\def\D{\Delta}
\def\tD{\tilde{\Delta}}
\def\cA{{\cal A}}
\def\cH{{\cal H}}
\def\e{\mbox{e}}

\begin{document}
\title{Landau-Zener transitions for Majorana fermions}
\author{Sergei Khlebnikov}
\email{skhleb@purdue.edu}
\affiliation{Department of Physics and Astronomy, Purdue University, 
West Lafayette, IN 47907, USA}
\begin{abstract}
One-dimensional systems obtained as low-energy limits of hybrid 
superconductor-topological insulator devices provide means of production, transport,
and destruction of Majorana bound states (MBSs) by variations of the magnetic flux.
When two or more pairs of MBSs are present in the intermediate state, there is 
a possibility of a Landau-Zener
transition, wherein even a slow variation of the flux leads to production of a
quasiparticle pair. We study numerically a version of this process, with  
four MBSs produced and subsequently destroyed, and find that, quite universally,
the probability of quasiparticle production in it is 50\%. This implies that 
the effect may be a limiting factor in applications requiring a high degree of quantum 
coherence.
\end{abstract}
\maketitle
\setcounter{footnote}{1}
Hybrid structures 
that consist of a strong three-dimensional topological insulator (TI) in contact with 
conventional ($s$-wave) 
superconductors have been predicted to host exotic excitations---Majorana
fermions and Majorana bound states (MBSs) 
\cite{Fu&Kane,Ioselevich&Feigelman,Cook&Franz,Ilan&al}. 
In particular, Fu and Kane \cite{Fu&Kane} have argued that
a TI-based Josephson junction (JJ) with a phase difference of $\pi$ hosts a longitudinally
propagating Majorana fermion and, if the phase varies gradually along the junction,
the fermion is trapped, near the point where the phase crosses $\pi$, into an MBS.
The trapping mechanism is that of Ref.~\cite{Jackiw&Rebbi}. If such a junction is
made a part of a superconducting interferometer (SQUID), the phase difference
can be controlled by the magnetic flux. As a result, it becomes possible to produce,
transport, and annihilate MBSs by varying the flux. Such a process has been considered
theoretically in Ref.~\cite{Potter&Fu}. 

\begin{figure}[b]
\begin{center}
\includegraphics[width=3.25in]{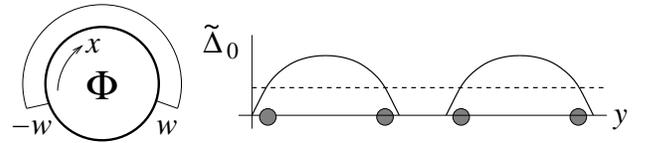}
\end{center}                                              
\caption{(Left) Cross section of a SC contact
deposited on the surface of a TI nanocylinder. The TI surface is shown by a thicker line.
(Right) Proximity-induced pairing amplitude modulus in the TI,
as a function of the
longitudinal coordinate, for the case when two such contacts are present. 
The dashed horizontal line represents the value $\delta(\Phi)$ (defined in the text)
for some longitudinal flux $\Phi$. As $\Phi$ changes, $\delta(\Phi)$ sweeps through 
the range of $\tD_0$.
The points where it intersects $\tD_0(y)$ are locations of MBSs (shaded circles).}
\label{fig:dev}  
\end{figure}

The mechanism of Ref.~\cite{Jackiw&Rebbi} 
as a means of producing and manipulating MBSs is quite generic. For example,
an MBSs can be trapped by a spatial variation of the modulus of the pairing
amplitude, rather than its phase. Under suitable conditions, that occurs, for
instance, in a device that consists of a number of 
superconducting (SC) contacts deposited along a TI nanocylinder or nanoprism
(Fig.~\ref{fig:dev}). Our interest in this system has been stimulated
by the experiments of Kayyalha {\em et al.} \cite{Kayyalha&al} on
nanostructures of this type.

Because this system 
presents a potentially useful alternative to a planar SQUID, 
and to underscore the generality of the trapping mechanism, let us briefly
describe the argument for the existence of MBSs in it (full details will be
published elsewhere \cite{SK:elsewhere}). The argument is 
based on the Hamiltonian of Fu and Kane \cite{Fu&Kane} adapted to a
cylindrical surface \cite{Zhang&al} and coupled to an external vector potential $\cA_x$.
 
Unlike Refs.~\cite{Cook&Franz,Ilan&al}, we do not assume that
the proximity-induced pairing amplitude $\D(x,y)$ in the Hamiltonian of the TI surface
states extends over the
entire circumference. Rather, each contact has a break, in which $\D(x, y) = 0$.
The break turns each contact into a quantum interferometer, which can be controlled by
the longitudinal magnetic flux $\Phi$. 

Suppose the SC layer is thick enough to lock the phase of
$\D$ to $\cA_x$, that is
\be
\D(x,y) = \tD(x,y) \e^{-2 ie \cA_x x} \, ,
\label{pair}
\ee
where $\tD \geq 0$ is real; $e > 0$ in the exponent is the magnitude of the electron charge. 
$\cA_x$ has been made $x$-independent 
by a gauge choice, so that $\Phi = \cA_x L_x$, where $L_x$ is the circumference. 
Note that, for a
general $\cA_x$, the state (\ref{pair}) is possible only
because the phase of $\D$ is not restricted by
periodicity in $x$. That would not be so in the absence of a break in the SC layer.

The simplest case is
when the chemical potential $\mu$ is a constant, while 
$\tD(x,y)$ is a slowly varying function of $y$, $\tD(x,y) = \tD_0(y)$, 
for $-w < x < w$ and zero otherwise (here we consider the version in which
$w$ is independent of $y$). In this case, we can at first neglect 
$\partial_y$ in the Fu-Kane Hamiltonian. We then find that, for any $\Phi$ in a certain
range, there is a value  $\tD_0= \delta(\Phi)$ 
at which the Hamiltonian has zero-energy states. 
With $\partial_y$ reinstated, intersections of $\tD_0(y)$ with $\delta(\Phi)$ 
become locations of MBSs (see Fig.~\ref{fig:dev}). 
As $\Phi$ changes between 0 and $\Phi_0 = \pi \hbar /e$ (the superconducting
flux quantum), these MBSs are produced, travel, and disappear, 
as the intersection points would.  
If $\tD_0(y)$ is not slowly varying, stability of an isolated MBS can
be argued by using a mod 2 index, similarly to the argument 
in the discussion of global anomaly in Ref.~\cite{Witten:1982}. 

In trapping by modulus, as in trapping by phase \cite{Fu&Kane}, 
the low-energy theory
of the MBSs is based on the one-dimensional (if we only
count the spatial dimensions) Dirac equation, with the additional ``reality'' 
restriction on the fermion. One can show that this low-energy theory
applies also to SC-semiconductor wire hybrids,
proposed as a platform for MBSs in Refs.~\cite{Lutchyn&al,Oreg&al}, 
so our results hold for that case as well. 
The MBSs are trapped by the mechanism of \cite{Jackiw&Rebbi}
on ``kinks''---simple zeroes of the fermion mass term.

Here, we would like to address one particular aspect of the physics of that
low-energy theory.
The energy of two well-separated MBSs is exponentially small for large 
separations. So, in
a long enough device, the process described above can be viewed as a passage through
an anti-crossing of two or more exponentially close energy levels. 
If the process is adiabatic 
on the scale of the typical quasiparticle energy but not on the scale of 
the exponentially small level spacing of the MBSs, one may ask if there is a version
of the Landau-Zener (LZ) 
effect, namely, a significant probability of transition from the ground to an excited 
state.

The difference between the present case and 
the standard LZ case \cite{LL} is two-fold. First, there is a difference in
the language of description. The Fermi statistics is best described by 
second quantization, so the natural language to discuss the process in our case
is that of particle production by a time-dependent external field. 
If the system starts in the ground state with no MBSs, then
goes through production, transport, and annihilation of those---so that none
are left in the final state---it 
will in general end in a linear combination of the ground state and states containing 
quasiparticles. Using the formalism of in- and out-states, we can write this as
\be
|0_{in} \rangle = f_0 |0_{out}\rangle + f_2 b^\dagger c^\dagger |0_{out}\rangle 
+ \ldots \, ,
\label{in}
\ee
where $b^\dagger$ and $c^\dagger$ are quasiparticle creation operators, and $f_0$,
$f_2$ are the transition amplitudes. 
Note that even
in the simplest case (when the additional terms indicated by $\ldots$ are absent), the
process requires participation of at least two quasiparticle modes (corresponding to
the two operators in (\ref{in})). A pair of MBSs gives rise to a single excitation
mode, so we conclude that the minimal setup in which a non-trivial effect is possible 
is that involving four MBSs.

The second difference between our problem and the original LZ calculation is that 
the latter applies to a specific form of anti-crossing of levels, 
namely, an avoided crossing of two straight lines. That form is generic when 
the off-diagonal elements of the $2\times 2$ reduced Hamiltonian are approximately
constant at the crossing point, while the diagonal terms have simple zeroes. 
In our case, however, the matrix elements do not have this structure:
if one were to neglect the overlap between 
the MBSs, their energies would be zero identically.

One may compare the effect described here to transitions induced by a change of
the phase difference between the ends of a proximity-coupled
wire that supports two MBSs at the ends \cite{Kitaev} or to LZ transitions induced by
a phase sweep in a wire that supports four MBSs---two at 
the ends and two more in the middle
\cite{San-Jose&al,Pikulin&Nazarov,Dominguez&al}. 
The difference is that, in either of those cases,
the MBSs exist permanently, while in our case
they are only present during the intermediate stages. Note in this connection
that the avoided crossing considered
in Refs.~\cite{San-Jose&al,Pikulin&Nazarov,Dominguez&al} is of the conventional 
straight-line type.

Here, we describe
a computation of the LZ effect---or, equivalently, particle 
production---in the setting corresponding to Fig.~\ref{fig:dev}, defined
in particular by the absence of MBSs in the initial
and final states and by exponential crossing of the levels.
We consider the minimal case identified above: two kink-anti-kink pairs, the
relative positions (and existence) of the kinks and antikinks being controlled by 
a single parameter, the separation $a$. The computation is done within the
4-dimensional subspace comprised of the first-quantized fermion states with 
energies closest to zero. We construct the $4\times 4$ $S$-matrix for these states
by numerically solving the evolution equations and then convert it into a
second-quantization relation of the form (\ref{in}). 

We find that a pair of quasiparticles is produced with probability 50\%, 
i.e., $f_2 = f_0 = 1/\sqrt{2}$ (with no additional terms present). This result 
has a curious degree
of universality: as far as we can tell, it applies regardless of the detailed
form of the wavefunctions, as long as the conditions of adiabaticity (as
defined above) and presence of well-separated MBSs during the intermediate stage
are intact. With these values of $f_0$ and $f_2$, Eq.~(\ref{in}) becomes equivalent 
to a fusion rule for non-Abelian anyons in the two-dimensional model of 
Ref.~\cite{Kitaev:2006}. That rule has been predicted to hold also for MBSs localized 
on triple junctions in a TI-based JJ circuit \cite{Fu&Kane}.\footnote{For that system,
derivation of the fusion rule can rely on the premise that the Majorana fermions
$\gamma_1,\ldots,\gamma_4$ are ``immutable,'' i.e., the same 
(apart from a single sign change \cite{Fu&Kane}) before and after
the transition; only their interactions change. That is not {\em a priori} obvious
in our case, as the MBSs are produced at one point and destroyed at another. 
The direct method described here is free of any such assumption.}
One implication of the universal nonzero value of $f_2$
is that any novel technology that aims to 
manipulate MBSs will have to contend with the possibility of quasiparticle
production and its associated effects (decoherence, heating, etc.).

Experimental verification of Eq.~(\ref{in}) for the system of Fig.~\ref{fig:dev}
is possible along the lines proposed for the planar circuit 
in Ref.~\cite{Fu&Kane} (and adapted to 
nanowires in Ref.~\cite{Alicea&al}). Namely, a measurement of the Josephson current 
between the contacts will produce one of two different values,
depending on which component of the linear superposition is projected out, with
50\% probability for each.

Our starting point then is the (1+1)-dimensional Dirac equation with 
a $y$-dependent mass ($y$ is the coordinate along the single spatial dimension):
\be
i \partial_t \Psi = [-i \sigma_y \partial_y + M(y) \sigma_z ]  \Psi 
\equiv \cH_2 \Psi \, .
\label{dirac}
\ee
Here $\Psi$ is a two-component fermion operator, and $\sigma$ are the Pauli matrices.
(We set both $\hbar$ and the velocity parameter to 1.)
The first-quantized Hamiltonian $\cH_2$ defined in (\ref{dirac}) has a particle-hole 
symmetry: if 
\be
V_n(y) = [A_n(y), B_n(y)]^T
\label{V}
\ee
is an eigenstate of $\cH_2$ with energy $E_n$, then
\be
V'_n(y) = [B_n(y), A_n(y)]^T
\label{V'}
\ee
is also an eigenstate, with energy $-E_n$. Because $\cH_2$ is real, we can 
choose both $A_n$ and $B_n$ to be real as well.

The Majorana case occurs when the components of $\Psi$ are constrained to be
Hermitian conjugate of each other:
\be
\Psi = (\psi , \psi^\dagger)^T \, .
\label{maj}
\ee
In this case, 
the operator solution to (\ref{dirac}) is
\be
\Psi(y,t) = {\sum_n}' [a_n V_n(y) e^{-iE_n t} + a_n^\dagger V_n'(y) e^{iE_n t}] \, ,
\label{sol}
\ee
where $a_n$, $a_n^\dagger$ are fermionic annihilation and creation operators, and
the prime on the sum indicates that each pair of states $V_n, V'_n$
related by the particle-hole
symmetry is included only once; in particular, $E_n \geq 0$.

A single simple zero of $M(y)$ (a ``kink'') in the infinite system would be a location of
a Jackiw-Rebbi bound state \cite{Jackiw&Rebbi}, for which $E_n = 0$ and 
$V_n = V_n'$ up to a sign. Substituting this into (\ref{sol}) we see 
that both components of $\Psi$ are expressed through a single Hermitian linear
combination of $a_n$ and $a_n^\dagger$, a Majorana fermion. Motivated by realistic 
examples, for which (\ref{dirac}) is the low-energy limit, we consider instead the
case when $M(y)$ has an even number of kinks, which can be produced and 
destroyed in pairs by variations of $M(y)$. In this case, there are no strictly 
vanishing $E_n$ but, when all the kinks are well separated, there are some 
exponentially small $E_n > 0$, one for each pair of kinks.

Next, we review the origin of LZ transitions. Suppose a first-quantized Hamiltonian
$\cH$ depends slowly on time, and there are two or more eigenstates $V_n(t)$ of
the instantaneous $\cH(t)$ whose energies, $E_n(t)$, undergo an avoided
crossing. To the extent we can restrict the time evolution to the 
subspace spanned by these eigenstates, we can search for a solution to the full 
time-dependent Schr\"{o}dinger equation (SE) $i \partial_t \Psi(t) = \cH(t) \Psi(t)$ in the
form
\be
\Psi(t) = \sum_n c_n(t) V_n(t) e^{-i \int^t E_n dt'}  \, .
\label{tdep}
\ee
Projecting the SE back onto the subspace, one obtains
evolution equations for the coefficients $c_n$. If, as in the present case,
the eigenfunctions are real, these equations simplify into
\be
\partial_t c_m +  \sum_{n\neq m} e^{i \int^t \Delta E_{mn} dt'} \Pi_{mn} (t) c_n  = 0 \, , 
\label{cm} 
\ee
where $\Delta E_{mn}(t) = E_m(t) - E_n(t)$, and
\be
\Pi_{mn} (t) = - \Pi_{nm}(t) = \langle V_m| \partial_t V_n \rangle \, .
\label{Pi}
\ee
Two limiting cases can be considered. When a $\Pi_{mn}(t)$ varies slowly compared to the
phase factor accompanying it in the equations, even a large $\Pi_{mn}(t)$ will be 
ineffective;
this is the adiabatic limit. In the opposite limit, of interest to us here, 
all $\Delta E_{mn}$
can be neglected during the entire time when $\Pi_{mn}$ are operational. Then, probabilities
of transition between different $V_n$ are
determined solely by $\Pi_{mn}(t)$. 

Note that in the latter limit, if $\cH$ depends on time
through a dependence on a parameter $a$, time can be removed from the equations
entirely, by replacing $\partial_t$ with $\partial_a$ in both (\ref{cm}) and
(\ref{Pi}). In other words, the precise form of the $a(t)$ dependence does not matter
(as long as the conditions allowing the use of the limit are satisfied).
In the original LZ case,
corresponding to an anti-crossing of two levels represented by straight lines,
the probability of the transition in this limit approaches unity \cite{LL}.
For the present case, it must be computed anew.

The case of four states (with real wavefunctions) involves four time-dependent
coefficients, $c_0$ through $c_3$, and six distinct matrix elements $\Pi_{mn}$
of the form (\ref{Pi}).
We consider four eigenstates of $\cH_2$, Eq.~(\ref{dirac}), with energies 
closest to zero and label them in the order of increasing energy, so that the 
particle-hole symmetry relates $V_0$ (the lowest-energy state of the four) to $V_3$, and
$V_1$ to $V_2$. In our earlier notation, this means that $V_0 = V_3'$ and $V_1 = V_2'$.
In the limit when all the energy differences are small (in the above-stated sense),
the evolution equations become
\be
\partial_t c_m + \sum_{n\neq m} \Pi_{mn}(t) c_n = 0 \, ,
\label{evol}
\ee
where $m,n=0,\ldots, 3$.

We have obtained the eigenstates $V_n$ required for computation of $\Pi_{mn}$ by
discretization of (\ref{dirac}) on a one-dimensional grid and
numerical diagonalization.
The fermions are assumed to obey antiperiodic boundary conditions on
the segment $0\leq y \leq L_y$.
In the continuum limit, we aim for $M(y) = \phi(y)$, where $\phi(y)$ is composed of
kinks and antikinks at positions $y_a = y_L - a$, $y_b = y_L + a$, $y_c=y_R - a$, and 
$y_d = y_R + a$, as follows:
\[
\phi(y) = \phi_0 \left[ -1 + K(y - y_a, \xi_L) - K(y - y_b, \xi_L) \right.
\]
\be
\left. {} + K(y - y_c, \xi_R) - K(y - y_d, \xi_R) \right] \, ,
\label{phi}
\ee
where 
\be
K(y, \xi) = \tanh (y / \xi) 
\label{kink}
\ee
is the kink, and $-K$ is the antikink; $\phi_0 > 0$ is a constant. 
(In the discrete version, it is convenient to deviate somewhat from $M(y) = \phi(y)$
 in order to better preserve the continuum form of 
the MBS wavefunctions.)

In each simulation, we pick specific values of the parameters 
$\phi_0, y_L, y_R, \xi_L, \xi_R$ and
vary $a$, the remaining parameter, which controls
the kink-antikink separation. When $a = 0$, we have $\phi = -\phi_0$, a
uniform state without kinks or MBSs. As $a$ increases from zero, a kink-antikink pair
appears at $y = y_L$ and another at $y = y_R$. At large enough $a$, two kinks 
disappear through the ends of the segment, and two annihilate in the middle.
So, they travel exactly as the intersection points in Fig.~\ref{fig:dev} do when the
horizontal line sweeps through the range of $\tD_0$ from top to bottom.
We choose $y_L$ and $y_R$ to be closer to the middle of the segment than 
to the ends, so the kinks heading towards the middle disappear first.

\begin{figure}
\begin{center}
\includegraphics[width=3.25in]{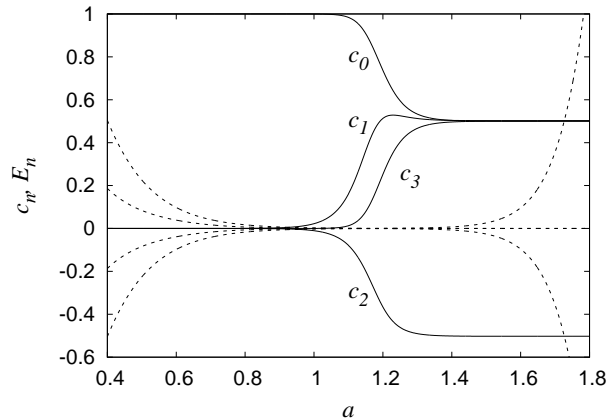}
\end{center}                                              
\caption{Evolution of the coefficients $c_n$ (solid lines),
obtained by numerical integration of 
Eq.~(\ref{evol}) for $\phi_0 = 5$, $y_L = 3$, $y_R = 7$, 
$\xi_L =0.4$, $\xi_R = 0.2$, $L_y = 10$, together with the corresponding eigenvalues of 
the instantaneous $\cH_2(a)$ (dashed lines).}
\label{fig:evol}  
\end{figure}

Fig.~\ref{fig:evol} shows the result
of a numerical integration of (\ref{evol}) with the initial 
condition $c_0 = 1$, $c_n = 0$ for $n > 0$ for a particular choice of the
parameters, together with the behavior of the four levels in question.
We see a nontrivial evolution of $c_n$ during
the time all the four levels are close to one another. Note that at the largest value
of $a$ shown, there are still two MBSs, corresponding to a kink and an antikink near the
ends of the interval. Their motion, however, does not produce any further change in $c_n$,
confirming the expectation that the minimum of four MBSs is required for a nontrivial 
effect. (That can also be verified in a simulation in which only two MBSs are produced
from the start.)

Let us combine the four $c_n$ into a column
\be
C = (c_0, c_1, c_2, c_3)^T 
\label{C}
\ee
and refer to the asymptotics of $C$ at small and large $a$ as $C_{in}$ and $C_{out}$,
respectively. We can then define a $4\times 4$ $S$-matrix connecting these asymptotics,
\be
C_{out} = \hat{S} C_{in} \, .
\label{Sdef}
\ee
Numerically, we find
\be
\hat{S} = \left( \begin{array}{cccc}
\half & -\half & \half & \half \\
\half & \half & \half & -\half \\
-\half & \half & \half & \half \\
\half & \half & -\half & \half 
\end{array} \right)
\label{Snum}
\ee
(For instance, the first column of this can be read off Fig.~\ref{fig:evol}.)
We find this result to be quite generic, i.e., equivalent $S$-matrices are obtained
for different values of the
parameters (provided four well-separated MBSs occur in the 
intermediate state).

To go over to second quantization, we replace the components of $C_{in}$ and 
$C_{out}$ with two sets of creation and annihilation operators, as follows:
\ba
C_{in} & = & (a_3^\dagger, a_2^\dagger, a_2, a_3)^T \, , \\
C_{out} & = & (b_3^\dagger, b_2^\dagger, b_2, b_3)^T \, .
\ea
Note that, as prescribed by (\ref{sol}), we associate annihilation operators with 
the positive energy exponentials, and creation operators with the negative 
energy ones.
Eq.~(\ref{Sdef}) becomes a Bogoliubov transformation relating the two sets.

Suppose the system is in the in-vacuum, defined by the conditions
\be
a_2 |0_{in} \rangle = a_3 |0_{in} \rangle = 0 \, .
\label{in-vac}
\ee
We wish to express $|0_{in} \rangle$ through the out-states, in particular, the
out-vacuum, defined by
\be
b_2 |0_{out} \rangle = b_3 |0_{out} \rangle = 0 \, .
\label{out-vac}
\ee
Inverting (\ref{Sdef}), we find
\ba
a_2 & = & \half (b_3^\dagger + b_2^\dagger + b_2 - b_3) \, , \\
a_3 & = & \half (b_3^\dagger - b_2^\dagger + b_2 + b_3) \, .
\label{BT}
\ea
The solution to (\ref{in-vac}) then is
\be
|0_{in} \rangle = \frac{1}{\sqrt{2}} (1 - b_2^\dagger b_3^\dagger) |0_{out} \rangle \, ,
\label{vac}
\ee
which shows that there is a 50\% probability to produce a quasiparticle pair.

The configuration (\ref{phi}) describes annihilation of
counter-propagating kinks that have been produced
(in pairs) at two different locations, $y_L$ and $y_R$. One may also consider 
the case when two pairs appear
at the same point (say, the middle of the interval) at different
times, and the kink from one pair eventually catches up with the antikink from
the other. (This can occur, for instance, for Josephson vortices
in a planar SQUID that contains a long junction of the type considered in
Ref. \cite{Potter&Fu}.) We have found that the $S$-matrix for that case 
is equivalent to (\ref{Snum}), so the probability of quasiparticle production 
is again 50\%.

\begin{acknowledgments}
I would like to thank Y. Chen, M. Kayyalha, and L. Rokhinson for an introduction
to their experimental work \cite{Kayyalha&al} and for discussions.
\end{acknowledgments}

\end{document}